\documentclass[preprint]{elsarticle}
\usepackage{lineno,hyperref}
\usepackage{amsmath}
\usepackage{xspace}
\usepackage{siunitx}
\usepackage{graphicx}
\usepackage{booktabs}
\usepackage{makecell}
\usepackage{subcaption}
\usepackage{color}
\usepackage{placeins}
\modulolinenumbers[5]
\journal{Nuclear Instruments and Methods A}
\newcommand{\microns}{\si{\micro}{m}\xspace}

\newcommand{\tritium}{$^{3}$H}
\newcommand{\carbon}{$^{14}$C }

\begin{document}
\begin{frontmatter}
\title{Medica-Plus: a Micromegas-based proof-of-concept detector for sub-becquerel tritium activity assessment at the service of oncological research}

\author{ 
F.\,Jambon, 
S.\,Aune, 
P.\,Baron, 
T.\,Benoit,
T.\,Bey,
D.\,Desforge,
E.\,Ferrer-Ribas,
A.\,Grabas,
M.\,Kebbiri,
I.\,Mandjavidze,
T.\,Papaevangelou,
M.\,Riallot,
M.\,Vandenbroucke}
\address{IRFU, CEA, Universit\'e Paris-Saclay, F-91191 Gif-sur-Yvette, France}

\author{ 
F.\,Beau,
V.\,Dive,
C.\,Malgorn}
\address{JOLIOT, CEA, Universit\'e Paris-Saclay, F-91191 Gif-sur-Yvette, France}
\author{
F.\,Malloggi,
A.\,Rousselot
}
\address{IRAMIS, CEA, Universit\'e Paris-Saclay, F-91191 Gif-sur-Yvette, France}
\author{
F.\,Carrel,
M.\,Trocmé}
\address{LIST, CEA, Universit\'e Paris-Saclay, F-91191 Gif-sur-Yvette, France}


\begin{abstract}
To fulfill needs in oncological research a new Micromegas detector has been developed to follow radiolabelled drugs in living organisms at the single cell level. This article describes the proof-of-concept of such a detector and compares its ability to detect and assess sub-becquerel \tritium~activities with a commercial $\beta$-imager. 
\end{abstract}

\end{frontmatter}
\newpage
\section{Introduction}
\label{section:Introduction}
Radio-labelling is one of the staple techniques to assess the behavior and biodistribution of drug candidates and their metabolites in pharmaceutical studies and development~\cite{Penner:2012}. Tritium (\tritium) and 14-Carbon (\carbon) radio-labelling are still among first-choice technologies for labelling active compounds, especially when the precise quantification of drug targeting and excretion are required~\cite{Isin:2012}. The field of oncology-related drug development is no exception to this rule, with maybe the additional constraint of tending towards lower doses due to the high toxicity of the drug itself~\cite{Lowenthal:1996}. Moreover, last decade discoveries in oncology highlighted the importance of cell heterogeneity on the drug affinity and metabolism~\cite{Marusyk:2012}, pushing further the need for precise quantification of drug, not only at the organ level, but also at the single-cell one. In this context, the  Medica-Plus project aims at developing a Micromegas-based~\cite{Giomataris:1995fq} $\beta$-imager detector to assess sub-becquerel activities in single cells. Gaseous detectors are particularly suited for $\beta$-imaging due to their low energy threshold, their good spatial resolution and their linearity. In addition they allow real-time imaging.

The Medica-Plus project involves a complex step of cell preparation in order to display unitary, well separated single-cells to be imaged in the detector that has been described elsewhere~\cite{Jambon:2020}. In that previous article we presented preliminary results obtained with a prototype detector measuring tritium-generated signal read globally over the detector active surface without any spatial information. In the present article, we report improvements in the measurement of sub-becquerel activities with an optimized detector, coupled with strip-reading electronics, and its comparison with a commercial $\beta$-imager performances. Further optimisations of the detector design for the $\beta$-imaging of single cells will be presented in a later article.

\section{Experimental set-up}
\label{section:Materials and Methods}

In this section the Micromegas detection system, the sample choice   and its manufacturing process and the activity assessment procedure are described in detail.
\subsection{Detection setup}
\FloatBarrier

\begin{figure}
    \centering
    \includegraphics[width=0.9\textwidth]{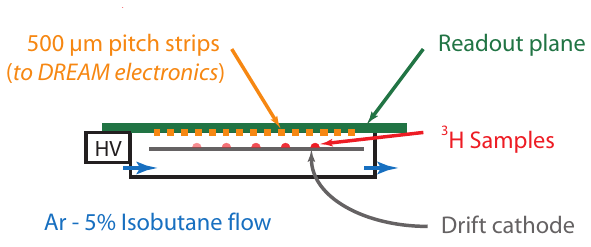}
    \caption{Sketch of the detection setup used.
    The Micromegas detector consists of an amplification gap of 128\,\microns and a drift gap of 5\,mm. The readout plane is made of 256 strips with a pitch of 500\,\microns.}
    \label{Detector}
\end{figure}

The detection system is based on the prototype detector that was described and characterized in detail here~\cite{Jambon:2020}. It consists of a Micromegas detector with an amplification gap of 128\,\microns, and a drift gap of 5\,mm. The readout plane is made of 256, 370\,\microns large copper unidirectional strips with a 500\,\microns pitch. The detection gas mixture is 95\% Argon - 5\% Isobutane flown continuously. 
The drift cathode is a thin film of 50\,nm-aluminized mylar from Goodfellow taut on a frame that adapts on drift supports allowing to maintain it in a given parallel position facing the reading plane. The frame is maintained with a nuts and bolts system enabling  for high voltage connection to the metallic face. This design was chosen so as to be able to minimize material waste when discarding the tritiated samples and the contaminated support to the appropriate waste management sector. In order to avoid readout plane contamination from samples that might drop on it, the detector is built so that the reading parts are always above the samples (Figure \ref{Detector}).

The detector has  been coupled with DREAM electronics~\cite{Acker2020}, allowing for independent strip reading. The maximum gain measured with a $^ {55}$Fe source is $\sim$17000~\cite{Jambon:2020}. The corresponding energy resolution (FWHM) is 29\%.

The DREAM electronics is used in self-trigger mode, meaning that each of the 256 strips is able to generate a hit signal for data acquisition when the charge variation over a defined amount of time exceeds a defined threshold value. The results shown in this article were obtained with a threshold of 54\,fC, to be compared to the dynamic range of 600\,fC used during these tests. 

\subsection{Tritium labelled samples}
Instead of living tumor cells, tritiated glucose deposits with decreasing activities were used. 
Particular attention was given to the solution preparation. To avoid any change in the self-absorption, the total glucose concentration in each solution was kept constant and equal to the content of the most active solution, by adding non-radioactive glucose. Solutions of tritiated and classical laboratory D-glucose were prepared in a solvent made of 50\% distilled water - 50\% ethanol (98\% purity, added for quicker drying time), with nominal activities of 0.185, 0.37, 1.85, 3.7, 18.5 and 37\,Bq/$\mathrm{\mu}$L. 

The activity of the solutions was assessed by liquid scintillation counting using 100\,$\mathrm{\mu}$L samples to reduce the measurement uncertainty. Triplicate 5-minute counting using a commercial scintillation counter were made for each solution activity. The best standard deviation achieved was about 3\%, which corresponds to the sum of systematic accuracy (one plastic tip is used per sample) and random error specification given by the pipette manufacturer. However the standard deviation could sometimes reach 15\%. The corresponding measurements and its associated true standard deviation ("sigma") obtained for each batch are given in Table \ref{GoutteSigmas}, in terms of activity per 1 $\mu$L droplet. The results will also be used as abscissa axis in the next section when comparing the measured activity with the deposited one.  

The samples were deposited at the same time on the Micromegas drift cathode and on the commercial $\beta$-imager microscope glass slides. Samples were then left to dry before closing the detector or placing the slides in the $\beta$-imager cassette. 

For the Micromegas detector, two sets of 1$\mu$L droplets of solutions of each activity were deposited on the drift cathode. The total activity obtained is thus 123\,Bq. For reference, the lowest activity corresponds approximately to the expected cell \tritium\, activity for the cell-measurement campaign. 
At the same time, droplets of the same solutions were deposited on a glass support. To assess a possible support material effect on the glucose crystallization process (discussed in Section \ref{AppDetect}), glass supports with a lining of the same aluminized mylar than the one used in the Micromegas detector have been produced. The same droplets deposition was made on both glass or mylar on glass supports and is as follows.

In order to assess the effect of the dynamic range on the beta-imager performance, two droplets of each solution in the 0.185 to 3.7\,Bq/$\mathrm{\mu}$L range were deposited on one glass support, and two droplets of the 1.85 to 37\,Bq/$\mathrm{\mu}$L range on another one. 

Three independent 30\,min long measurements were performed with all six sample activities in the detectors in order to assess the accuracy, but also the standard deviation of the measurement for each detector. For the $\beta$-imager, another series of measurements was made only with the “lower activity range" samples inside (referred later as “Reduced" range, in opposition to “Full" range).
The activity assessment methodology for each detector is described in the next section.

\subsection{Activity analysis procedure}
In this section, the activity assessment procedures for both detectors is detailed.
\subsubsection{Micromegas data analysis}
\begin{figure}
    \centering
    \includegraphics[width=0.75\textwidth]{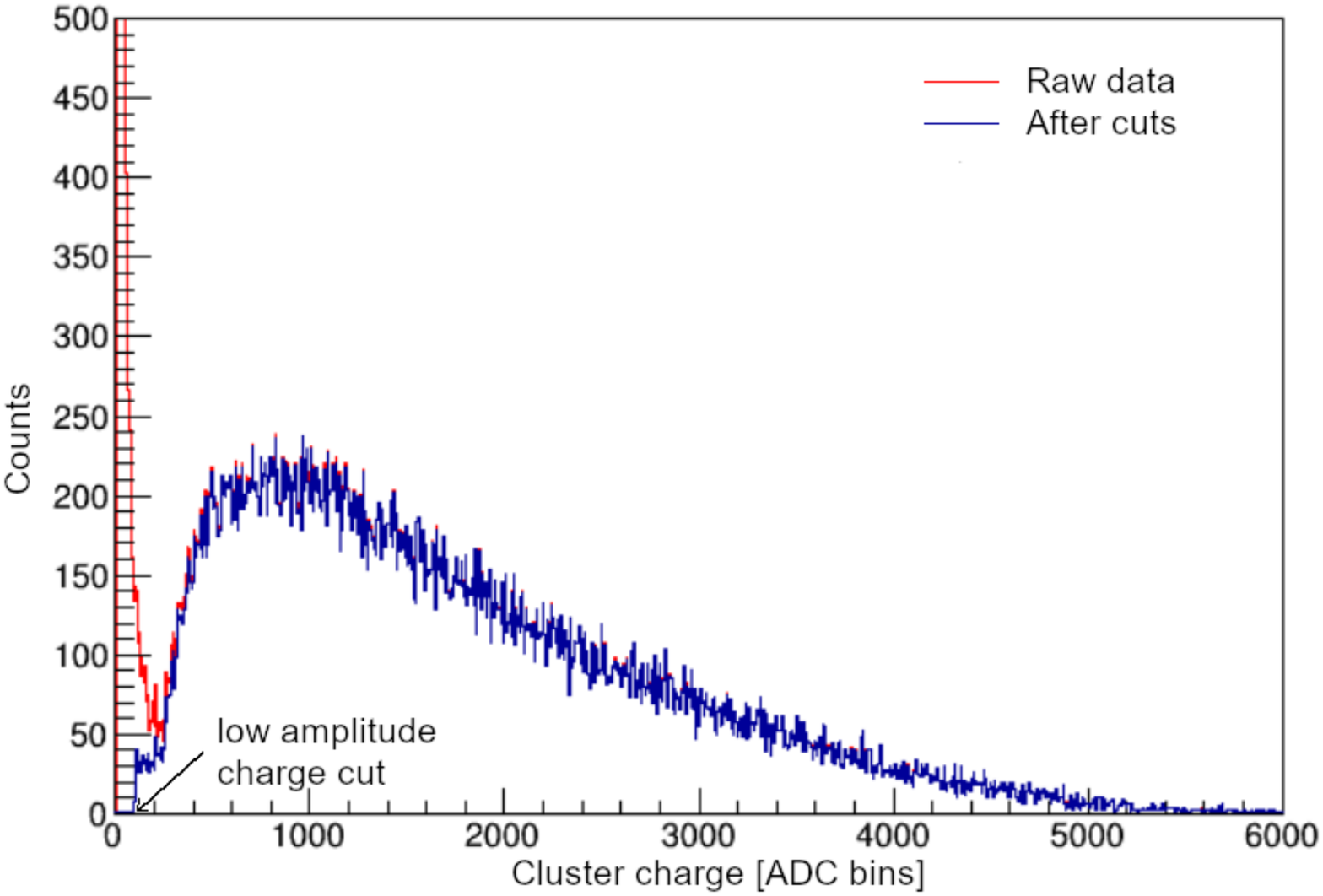}
    \caption{Effect of the data processing on the cluster charge distribution signal produced by a 30-minutes long measurement of \tritium-glucose samples;  the $\beta$ spectrum shape of the \tritium~energy distribution is recognizable.}
    \label{EnergyCuts}
\end{figure}

\begin{figure}
    \centering
    \includegraphics[width=0.495\textwidth]{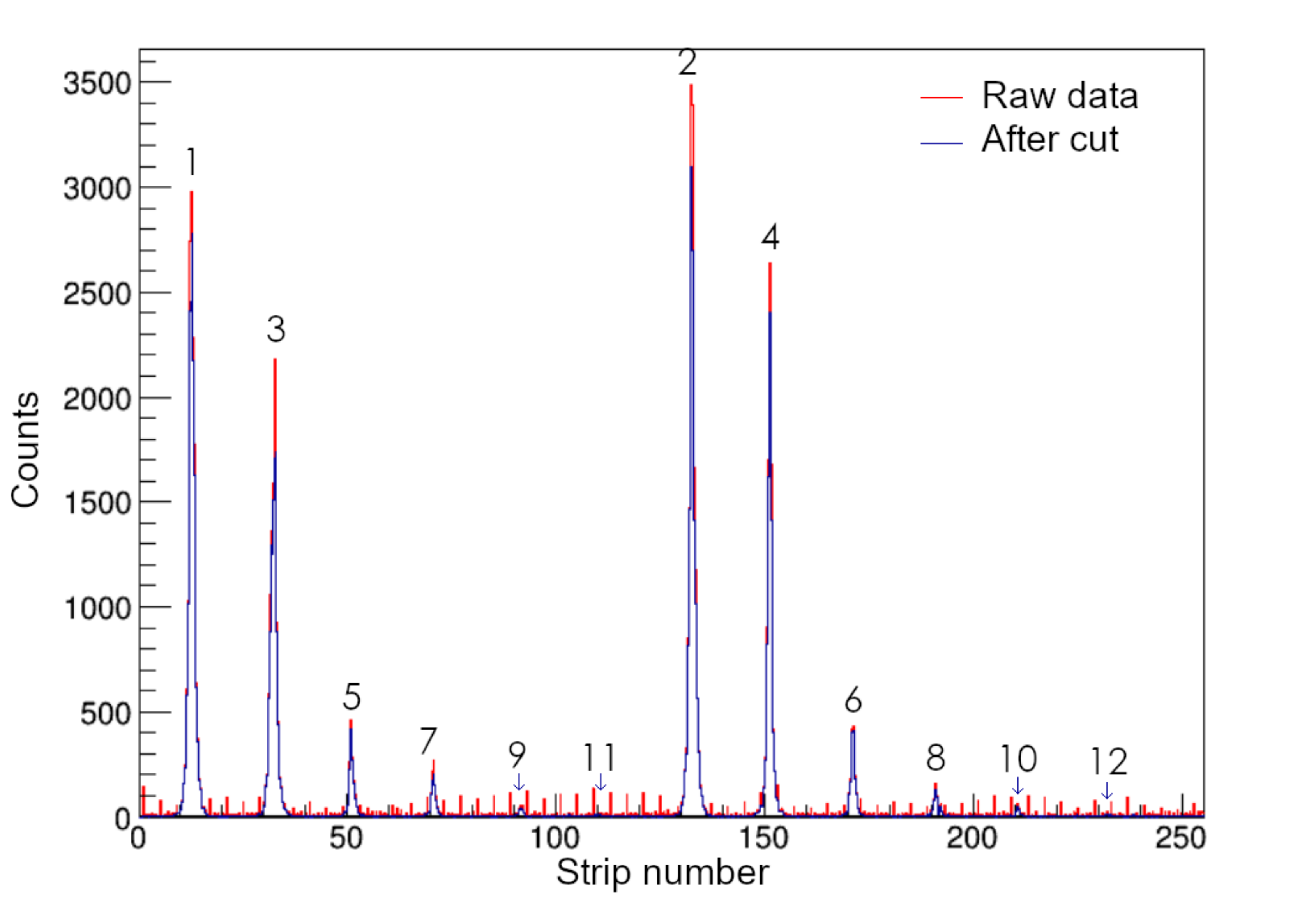}
     \includegraphics[width=0.495\textwidth]{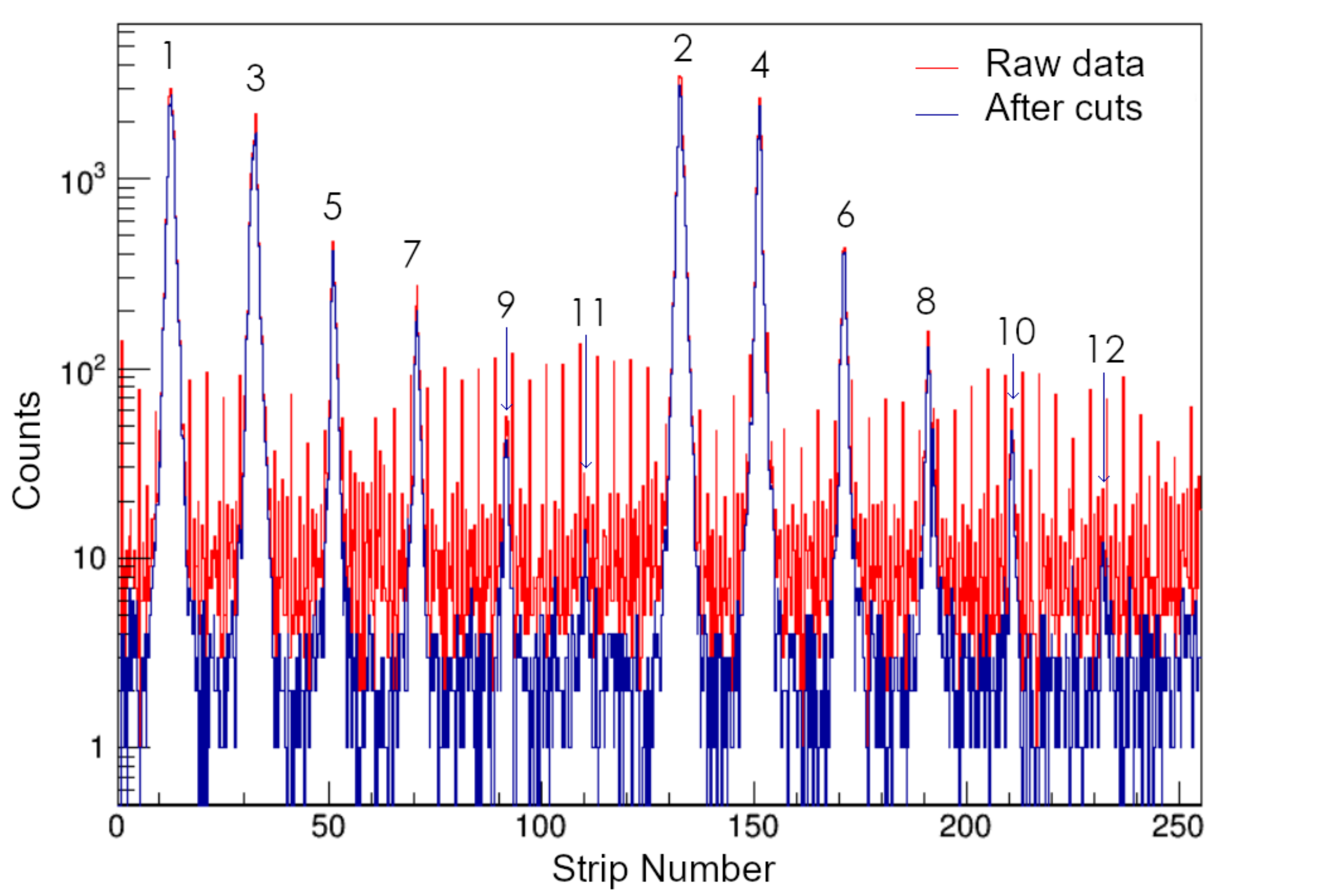}
    \caption{Reconstructed cluster position distribution produced by the \tritium-glucose samples during a 30-minute acquisition. Effect of the data processing on the result obtained.  Left: linear count scale; Right: logarithmic count scale.}
    \label{Cuts}
\end{figure}

\FloatBarrier
For the Micromegas measurement, all events triggering the DREAM electronics are recorded, and data is processed afterwards. First, pedestals are calculated and  then subtracted for each individual strip. Common noise subtraction is performed by groups of 32 strips. Then, events of significant charge collected on neighbouring strips during a hit window are gathered together in clusters. In this case, one cluster corresponds to one decay. 
A weighted arithmetic mean is computed on the distribution of charge amplitudes on each cluster strip, so that the cluster centre position could be assessed. Electronics noise is removed by a cut on low-amplitude clusters. This cut suppresses clusters with a maximum strip  charge lower than to twice the electronics threshold. Finally, a topological analysis is performed in order to keep only the clustered events compatible with \tritium-generated signals. Best results are obtained with a combination of two conditions: first, a minimum threshold is required on the energy of the larger contributing strip in a cluster; this discards large clusters  with low energy, mainly coming from noisy background that would escape the common noise suppression. Second, only the signal collected in the DREAM electronics between 2 and 7 samples time stamps is kept (one sampling interval being approximately 48 ns long).  The overall effect of the data processing on the clusters charge distribution is shown in Figure~\ref{EnergyCuts} and \ref{Cuts}, for a 30-minute long acquisition of \tritium-glucose samples in the Micromegas detector. 

The activity of each \tritium-glucose sample is assessed by counting the number of clusters per peak, divided by the run length to obtain an activity in becquerel. A simple routine that fits a linear background over the whole distribution and detects the peaks from their local maximum has been developed. Each peak is fitted  with a Gaussian function. The adjusted mean position and sigma are used to calculate the number of clusters retained per peak. Cluster count per peak is the total bin content of the cluster centers distribution between the Gaussian fit mean parameter, plus or minus, five adjusted sigmas. 
A background subtraction is then performed using the calculated local background at mean peak position, considered constant along the peak width. The background before cuts is constant at 9 counts (5.0\,mBq) per bin in raw data, and 5 counts (2.7\,mBq) per bin for a 1024-bins (for 256 strips) histogram. 

 The effect of this data treatment on the activity assessment is quantified on the right hand side of Table \ref{GoutteSigmas}, where the results obtained with the $\beta$-imager (and detailed in Section \ref{betaiprocess}) for both sample activity range conditions are also displayed. The activity determined without cuts or correction for each cluster peak is found in the “Raw" column; the “Cuts" column gives the activity after charge and topological cuts, while in the “Corrected" one, background subtraction is also applied. The peak numbers refer to those in  Figure \ref{Cuts} for Micromegas and Figure \ref{BetaTreatment} for the $\beta$-imager, while the “Scintillation counting" gives the expected activity of the 1\,$\mu$L sample solution. Finally, “sigma" columns stand for the standard deviation obtained over the series of three measurements. 
 
\begin{table}[!h]
    \centering
    \footnotesize
    \begin{tabular}{|c|c|c|c|c|c|c|c|c|c|c|}
        \hline
        Peak  & \multicolumn{2}{c|}{Scintillation}  & \multicolumn{4}{c|}{$\beta$-imager Corrected}  & \multicolumn{4}{c|}{Micromegas}\\
          Number & counting & sigma & Full& sigma & Reduced & sigma & Raw & Cuts & Corrected & sigma\\
         \hline
        1 & 36 & 6 & 10 & 1 & - & - & 11.75 & 11.55 & 11.53 & 0.20 \\
        2 & 36 & 6 & 9 & 1 & - & - & 11.57 & 11.30 & 11.28 & 0.17 \\
        3 & 21 & 4 & 7 & 1 & - & -  & 7.20 & 6.96 & 6.95 & 0.09 \\
        4 & 21 & 4 & 4 & 1 & - & - & 7.57 & 7.30 & 7.28 & 0.03 \\
        5 & 3.3 & 0.4 & 1.5 & 0.5 & - & - & 1.49 & 1.28 & 1.26 & 0.04 \\
        5bis & 3.3 & 0.4 & 1.7 & 0.5 & 1.2 & 0.04 & - & - & - & -\\
        6 & 3.3 & 0.4 & 1.5 & 0.5 & - & - & 1.58 & 1.36 & 1.34 & 0.02 \\
        6bis & 3.3 & 0.4 & 1.6 & 0.5 & 0.9 & 0.04 &  - & - & - & -\\
        7 & 1.48 & 0.08 & 0.7 & 0.5 & - & - &  0.97 & 0.71 & 0.70 & 0.03 \\
        7bis & 1.48 & 0.08 & 0.7 & 0.5 & 0.53 & 0.02 & - & - & - & -\\
        8 & 1.48 & 0.08 & 0.6 & 0.5 & - & - &  0.70* &  0.43* & 0.41* & 0.07 \\
        8bis & 1.48 & 0.08 & 0.7 & 0.5 & 0.53 & 0.02 & - & - & - & -\\
        9 & 0.272 & 0.006 & 0.1 & 0.3 & 0.12 & 0.01 & 0.493 & 0.186 & 0.169 & 0.009  \\
        10 & 0.272 & 0.006 & 0.1 & 0.3 & 0.11  & 0.01 & 0.531 & 0.201 & 0.179 & 0.006 \\
        11 & 0.13 & 0.02 & 0.1 & 0.2 & 0.070  & 0.004 & 0.435 & 0.128 & 0.106 & 0.020 \\
        12 & 0.13 & 0.02 & 0.1 & 0.2 &  0.061  & 0.004 & 0.348 & 0.105 & 0.086 & 0.024 \\
        \hline
       
        \hline
    \end{tabular}
    \caption{Effect of data treatment and background subtraction on data. Activities are in Bq. The significant digits are calculated over the series of three measurements of the same samples and correspond to the sigma column at the right of the activity for each measurement type, and to the error bars shown latter in plots.}
* \tiny{dead strip in the middle of the peak on the Micromegas detector. Not used for performance assessment}
\label{GoutteSigmas}
\end{table}

\subsubsection{$\beta$-imager activity analysis}\label{betaiprocess}
\begin{figure}
    \centering
    \includegraphics[width=0.75\textwidth]{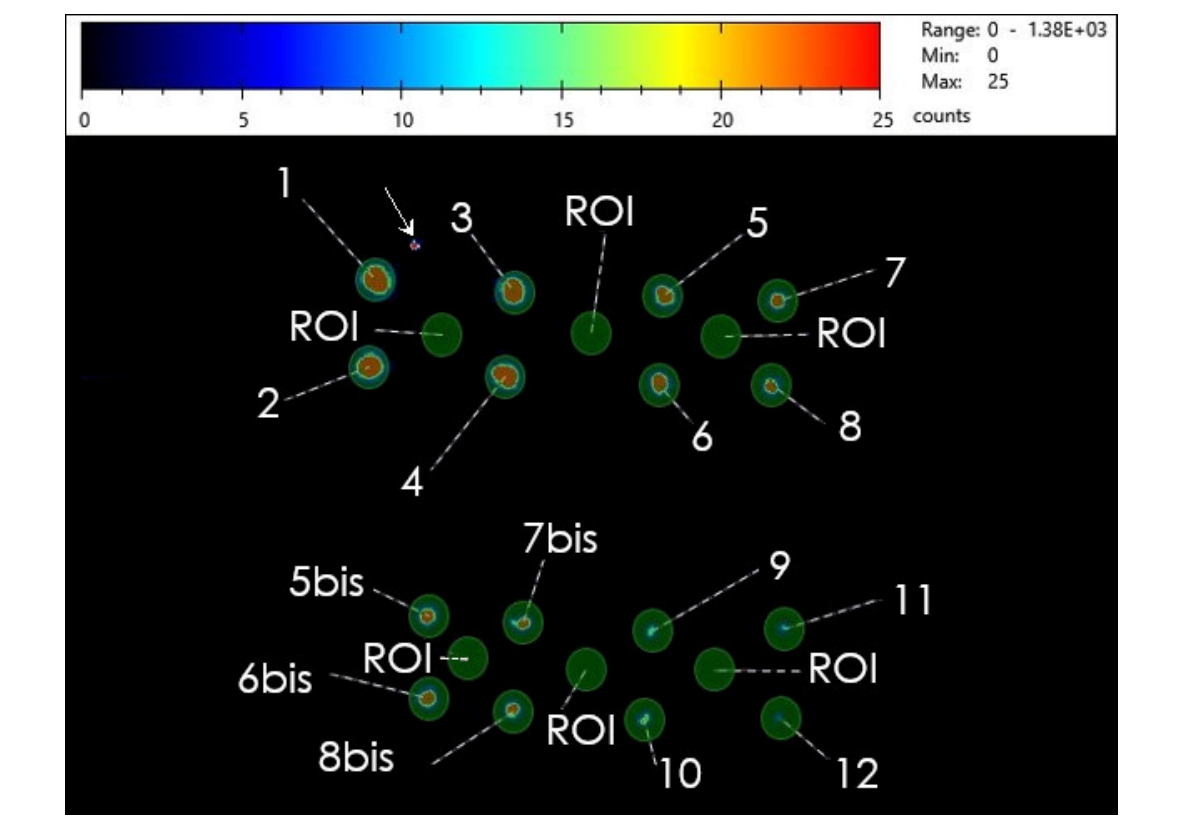}
    \caption{Illustration of the $\beta$-imager activity assessment analysis. Spots are numbered according to the activities of peaks in Figure 2 for consistency. ROI stands for Region Of Interest chosen for background estimation. The arrow (close to spots 1 and 3) points a hotspot, that might be a contamination on the support of the grid. For the Reduced Range measurement, only the bottom spots were counted.}
    \label{BetaTreatment}
\end{figure}

For the $\beta$-imager, a commercial Biospace Beta Imager 2000 (now renamed Biospace tRACER) was used. This imager is also a gaseous detector implementing a multi wire proportional chamber  with an optical readout. For each measurement, the activity per sample is assessed as follows. First, the color scale is adjusted so that the sample contours are highly visible. An illustration of the process is given on Figure~ \ref{BetaTreatment}; note that the spatial distribution is not reflected anymore by the color scale so that the samples look very different in span.
Then, samples are surrounded with a region of interest (ROI) where the total count of decay per mm² is calculated automatically by the $\beta$-imager software. Background noise is subtracted after choosing a similar ROI aside from the samples allowing the calculation of noise expressed in counts per mm$^2$. The corrected sample ROI value is then re-expressed in units of total counts, which are divided by the analysis time to obtain an activity in Bq. 
The results of activity assessment by the $\beta$-imager for both Full or Reduced (lower) activity range are given in Table \ref{GoutteSigmas}. Since the background level for the chosen ROI is about 6\,mBq, that is much smaller than the significant digits associated with the measurement, the detail of background subtraction effect is not given in Table~\ref{GoutteSigmas}.

\section{Results and Discussion}
\label{section:Results}
In this section, a comparison of the overall activity assessment obtained with the Micromegas detector and with the commercial $\beta$-imager is presented. Then background correction and detection efficiency is discussed. Improvement perspectives are presented in the last subsection.

\subsection{Comparison of \tritium~activity assessment performance}
 In order to assess the measurement stability and reproducibility and separate it as much as possible from the sample-to-sample variation for a same tritiated solution batch, a two-step process has been applied. First, for each sample/deposit, the mean activity and the associated standard deviation over three independent 30\,min measurements are calculated. Then a mean of the means, representing  the activity accuracy and its combined error are calculated. The error  represents a measurement-to-measurement deviation and thus an indicator of reproducibility quality. 
 \begin{figure}
    \centering
    \includegraphics[width=0.49\textwidth]{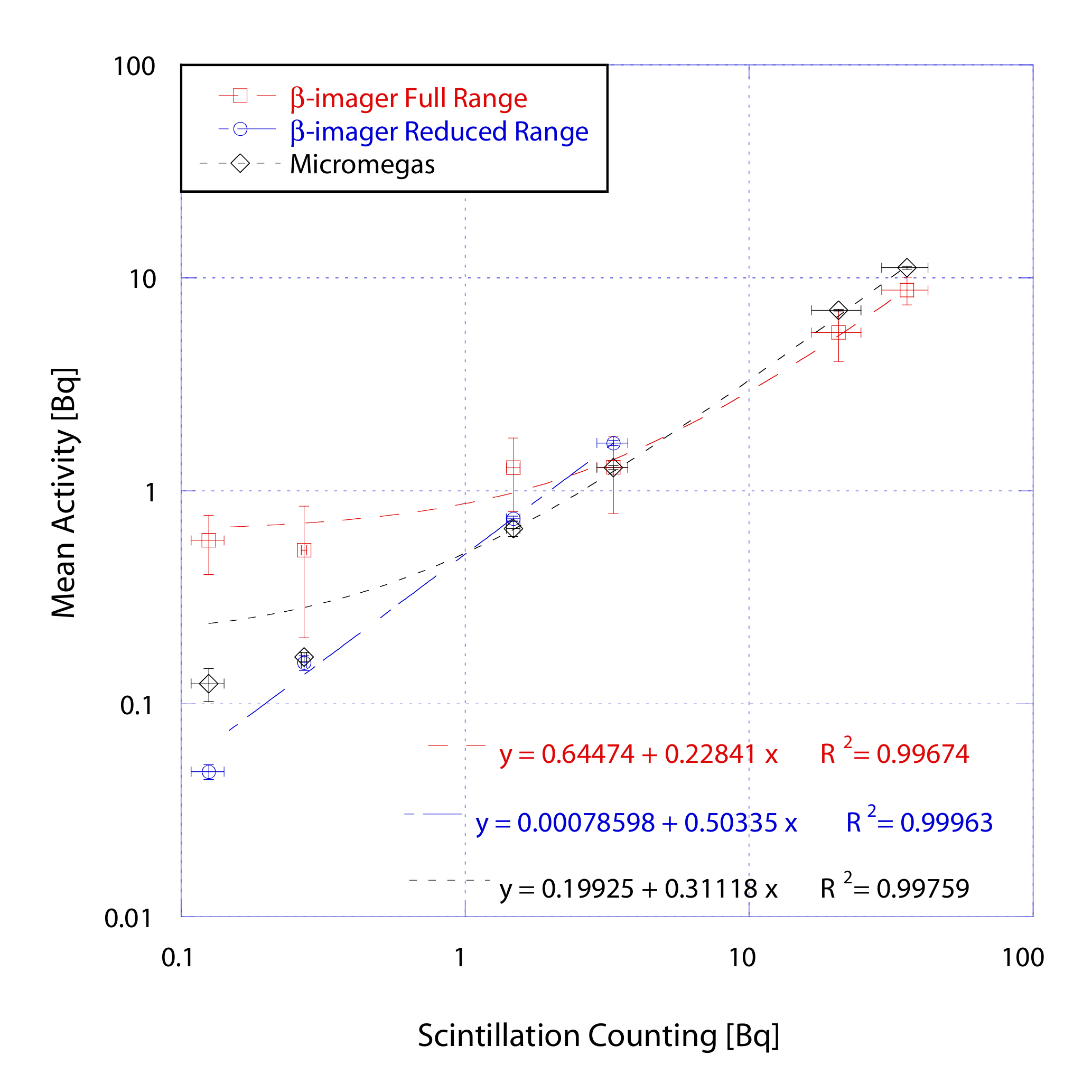}
    \includegraphics[width=0.49\textwidth]{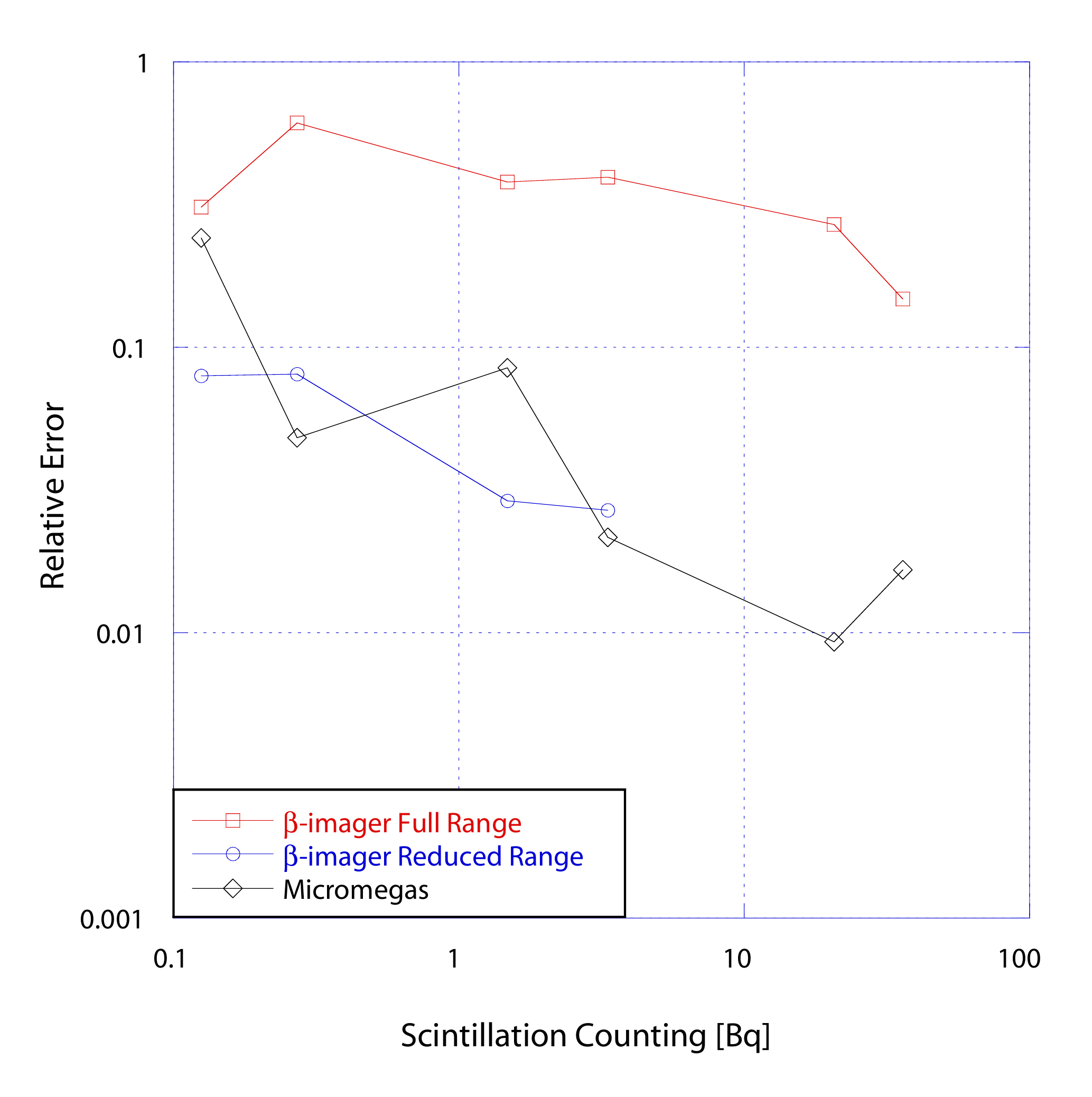}
    \caption{Right: Comparison of mean activity assessment (symbols) by a commercial $\beta$-imager for two dynamic ranges and by the Micromegas detector, linear regressions associated (lines and equations). Left: Relative error made by the detectors in different conditions versus sample activity }
    \label{TS6}
\end{figure}

The activities thus obtained are plotted versus the sample activity derived from the  scintillation counting of the \tritium-glucose solution batch, and displayed in Figure \ref{TS6}. A comparison of the results obtained with the full dynamic range of samples with the Micromegas (black lozenges) and the commercial $\beta$-imager (red squares) as well as the recounting of the Reduced (bottom half) range of activities with the $\beta$-imager (blue circles) is displayed. The vertical error bars (see “sigma"s in Table \ref{GoutteSigmas}) are the two-step standard deviation calculated as described above, while the horizontal ones correspond to the scintillation counting standard deviation. 
A linear regression is also plotted and the fitted parameters and associated R$^2$ are displayed in the figure. 
The slope is a measurement of the apparent detection efficiency while the constant parameter is an evaluative of the theoretical detection limit. In order to help evaluate the differences between each detector and conditions, a graph of the relative error, that is the ratio of the standard deviation over the mean value, is plotted Figure \ref{TS6} left.


First, comparing the activity assessment through in broad outline, and from the linear regression from Figure \ref{TS6}, it is clear that, overall,  for the full range of sample activity, the Micromegas-based detector (black curve, lozenges) does better than its commercial counterpart (red curve, squares) when used in the Full range mode. Indeed, the Micromegas detector data is associated with a better detection yield, linearity and projected detection limit. Also, the relative error associated with the activity assessment is better with the Micromegas detector: one to two orders of magnitude compared to one of the commercial $\beta$-imager with Full range mode (Figure \ref{TS6} left). 

\subsection{Background noise and optimal detection strategies}

In the case of the Micromegas detector, since each cluster is treated as a decay, the detector is used in counting mode. The gain setup chosen for optimal performance remains fixed so that the detection of a given event does only depend on its energy. 

\begin{figure}
    \centering
    \includegraphics[width=0.65\textwidth]{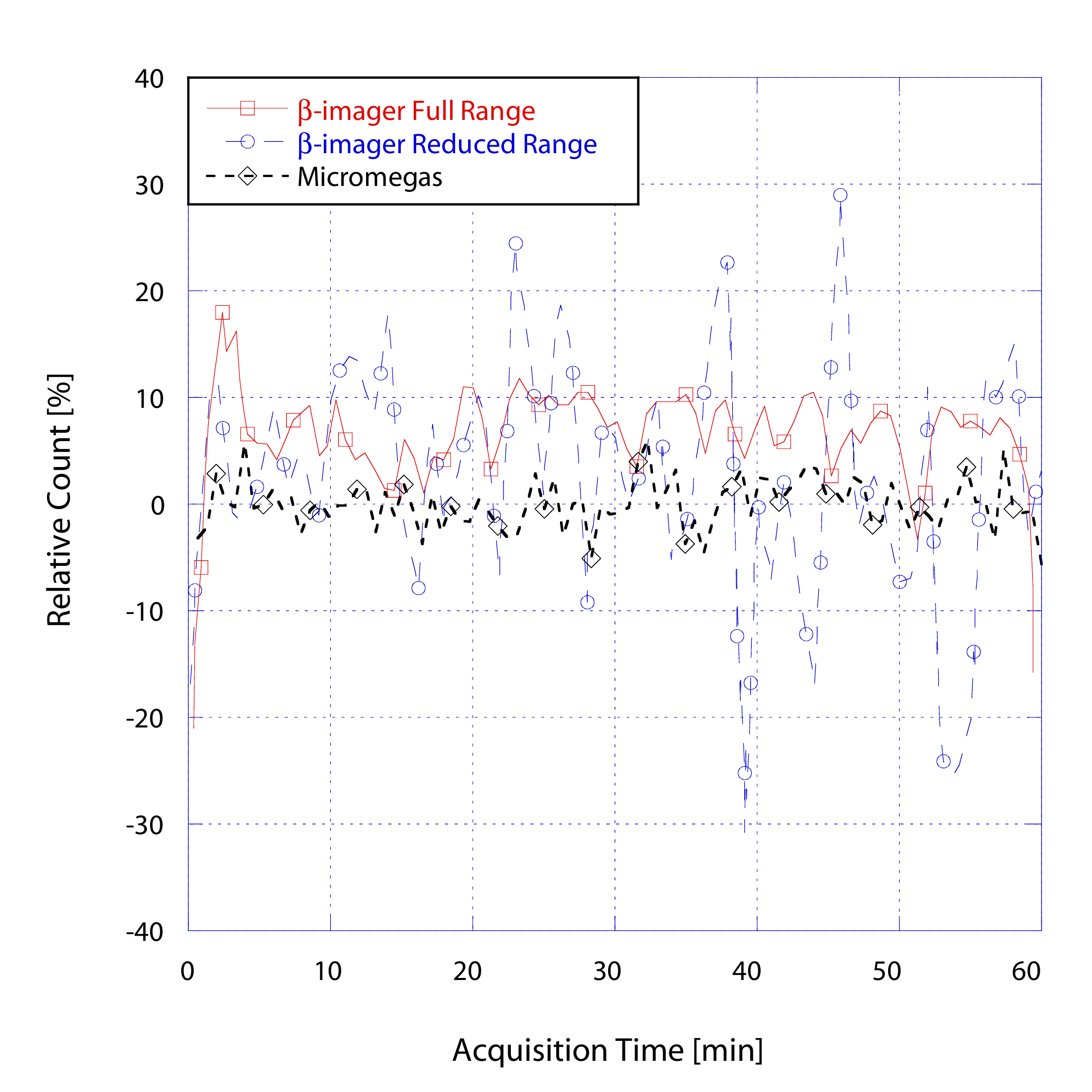}
    \caption{Relative detector count rates for a 1-hour acquisition : comparison between $\beta$-imager Full (red squares) and Reduced (blue circles) dynamic ranges and Micromegas (black lozenges)}
    \label{Oscillation}
\end{figure}

This is different from how the commercial $\beta$-imager works. Indeed, it adjusts its gain during the acquisition by modulating the high voltage applied internally. At low activities this results in an oscillation of its background level during the acquisition time, and eventually in sparks on the sample that can induce errors or in the worst case detrimental effects on the detector itself including contamination. 
An illustration of this oscillation process is given in Figure \ref{Oscillation} where the relative count rate around average for a one-hour acquisition has been plotted for both detectors.

\FloatBarrier

For this reason there is a strong effect of the dynamic range on the activity assessment, especially at low activity. As a result, the situation where spot-like samples have a broad range of activities in unfavourable for the commercial $\beta$-imager, while the Micromegas detector offers a more reproducible and stable measurement even for short acquisitions times.

Going back to Figure \ref{TS6} right, one can consider that the projected detection limit (the linear regression's constant parameter) is also an indication of how well the background is dealt with. 

Considering the background rate roughly constant and resulting essentially from the interaction of cosmic particles with the detector, it is a good strategy to correct the measurement with an average of the background. However, this induces a disadvantage in situations where the signal-to-background ratio is low (in the present case, this means at low samples activity). As a result, the commercial $\beta$-imager is disadvantaged when measuring the Full range of samples activities: as explained, its gain is set “low" and the signal-to-background ratio is lower than the best achievable one, because low energy clusters associated with \tritium~decay are not detected, whereas the cosmic particles-related tracks are. As a consequence, the error, especially on small activity samples, is high. On the other hand, for the Reduced (and lower) range of activities, the commercial imager sets a high gain and gets a better signal-to-background ratio, and does better than the Micromegas. 

\subsection{Apparent efficiency of the detectors} \label{AppDetect}

\begin{figure}
    \centering
    \includegraphics[width=0.7\textwidth]{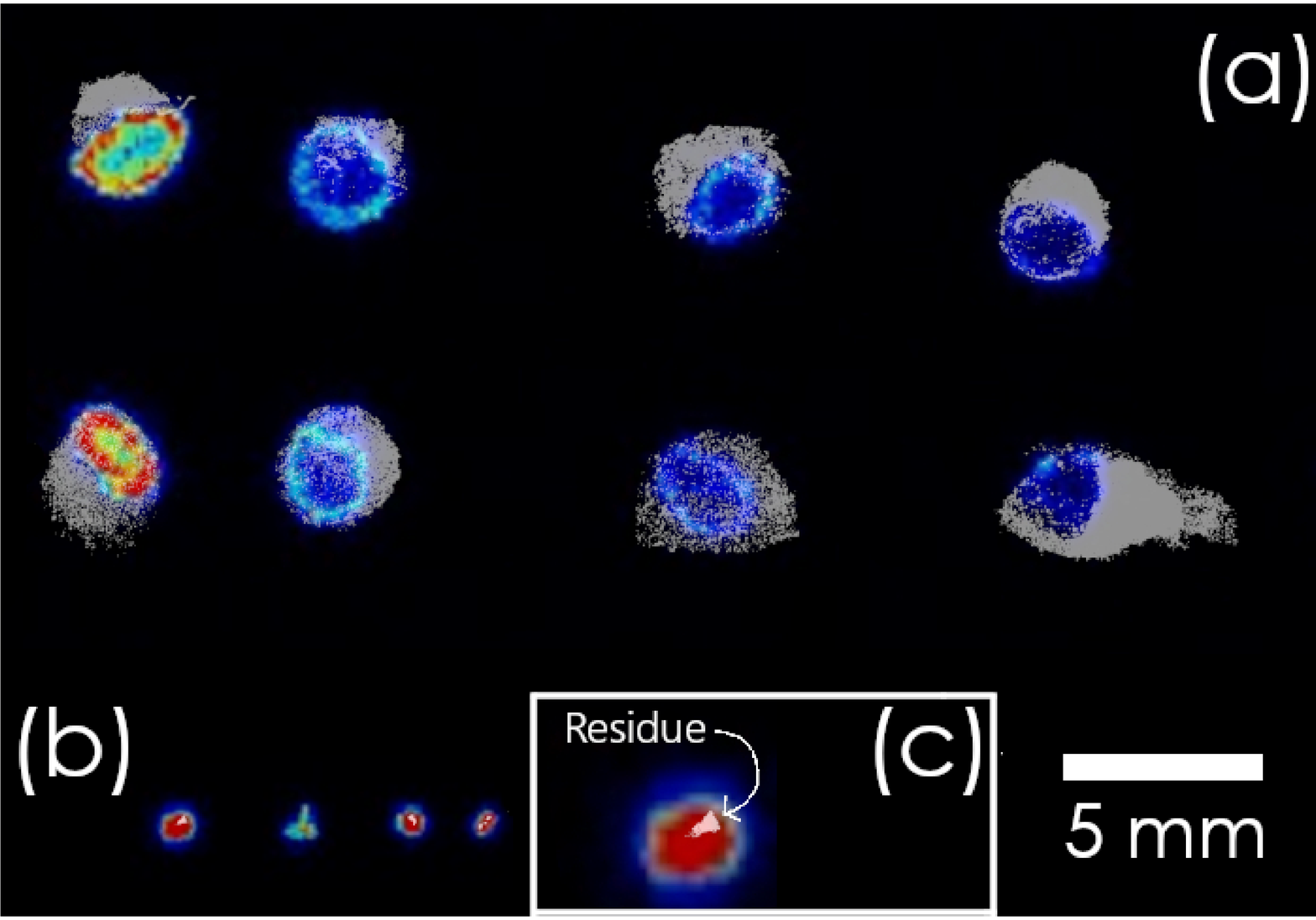}
  \caption{Distribution on glass (a) versus mylar  (b) support of crystallized matter (grey) versus radioactivity (color) seen in the commercial $\beta$-imager. Insert (c) : a zoom of the most active sample on mylar. Four sample activities are tested, one per column; the activity ratio from right to  left is 1:5}
   \label{LameMylar}
\end{figure}

The slope of the linear regression in Figure \ref{TS6} for the full activity range gives a detection efficiency  of 23\% and 31\% for the $\beta$ imager and the Micromegas respectively. This can be partly explained by the loss of solid angle, as only $\beta$s facing the conversion space of the detector and in the surface of the spot will reach the gas. 
It  might be that the solvent impurities, which are roughly constant in each droplet, contribute predominantly to the self-absorption compared to the glucose content in all conditions tested.
An illustration of this showed up during the fine-tuning of the measuring method offered here. Initially, for $\beta$-imager acquisitions the \tritium-droplets were deposited on glass only, but a strong difference in the apparent efficiency of both detectors appeared. It seemed correlated to a difference in the drying process. 
After the solvent has dried, on glass one could see clearly a white deposit on the glass, corresponding roughly to the space occupied by the initial 1\,$\mathrm{\mu}$L drop. By segmenting a picture of the microscope glass slice and superimposing the image of the tritium activity measured in the $\beta$-imager, Figure \ref{LameMylar}(a) is obtained (the color scale is there for illustration and is proportional to the tritium activity in a linear way; the dynamic range has 1:5 proportions). The same work was made on aluminized mylar - this time localizing the dry deposit through a microscope observation (Figure~\ref{LameMylar}((b)). Only a tiny spot was found (a zoom of the most active spot on mylar is given in the frame (c) for  better visualization). 

These observations highlight a strong support material effect on the distribution of both white deposit and radioactivity (that is, \tritium-glucose). On glass, there is a preferential crystallization of impurities as the solvent evaporates, then the glucose dries following a well-known coffee stain pattern, whereas on mylar both glucose and impurities crystallize at the same spot. In this configuration, the self-absorption phenomena is driven by the total impurity content for the mylar-supported samples, but only by the glucose content for the glass-supported samples.

In the series of tests presented in the previous section, the apparent efficiency is  31\% on mylar (Micromegas) or 23\% on mylar-on-glass ($\beta$-imager). The efficiency on mylar on glass is only higher (50\%) in the case of reduced dynamic range, for which the $\beta$-imager actually sets up a higher high voltage between sample and grids. 

This added to a likely combination of good background estimation and subtraction 
may explain this difference.

Figure \ref{LameMylar}(a) also demonstrates graphically a strong sample-to-sample variation in the local distribution of the activity. This dependence for the  crystallization pattern is one of the motives for way the two-step mean and standard deviations described in the previous section. 

To summarize, working with \tritium-glucose is a handy, quick and easy way to develop the Micromegas-based detector proof-of-concept and show its ability to detect and assess low \tritium\, activities. But the crystallization process sets up limitations that cannot be fully overcome nor compensated by a simulation work for instance.

\subsection{Perspectives for improvement}

\begin{figure}
    \centering
    \includegraphics[width=0.7\textwidth]{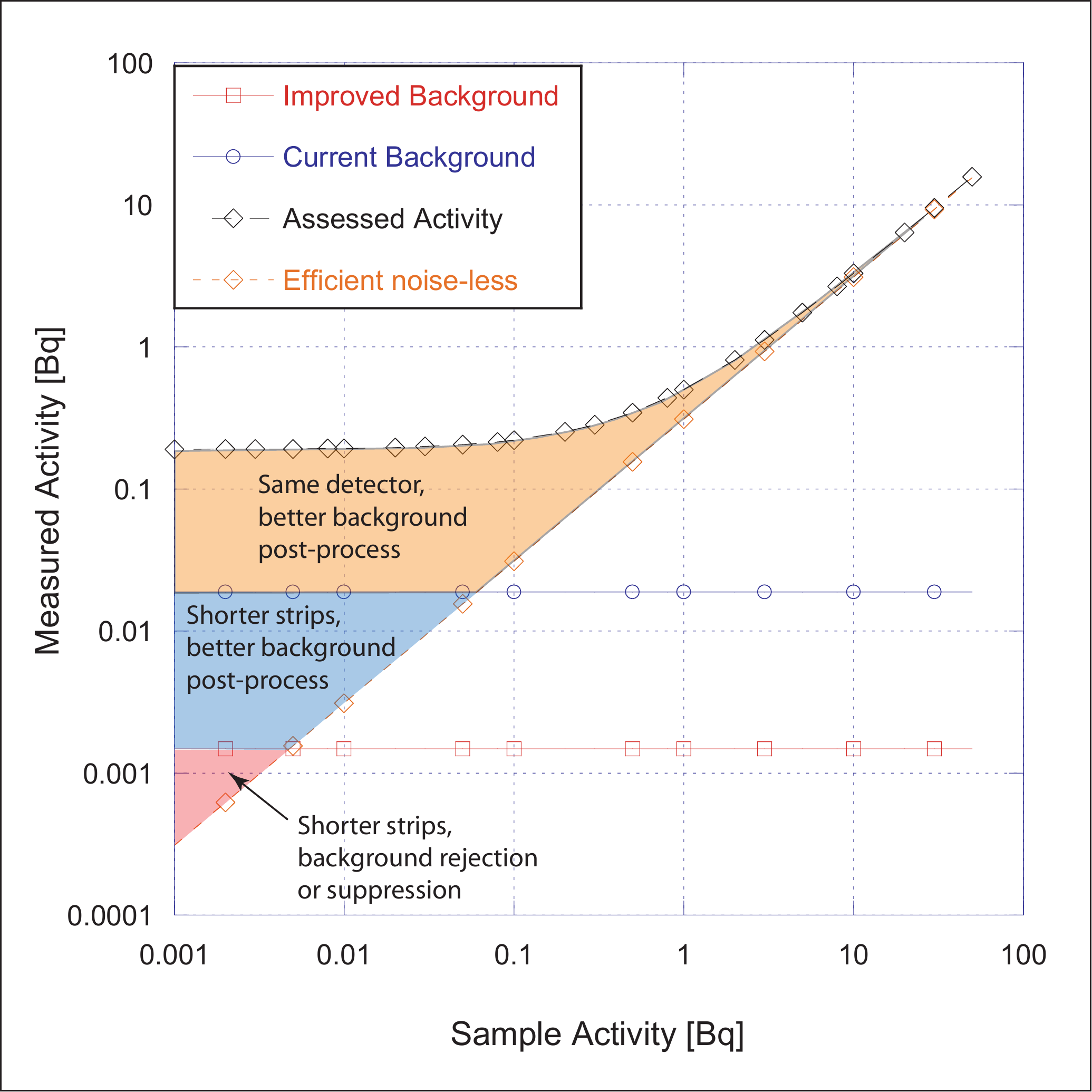}
    \caption{Domains of improvement considering the current activity assessment function of the Micromegas versus a potential perfectly noise-less detector with the same efficiency.}
    \label{Improvement}
\end{figure}

For the Micromegas-based detector, several domains of improvement exist, that can be attained by successive steps of background treatment, detector optimization, cosmics rejection. They are represented in Figure \ref{Improvement}, where the “Assessed Activity" curve is plotted according to the Micromegas linear regression fit from the previous section. Considering that the Micromegas detection efficiency is constant, improvement consists in getting closer to the straight “Efficient, noise-less" line that has the same slope, but no constant parameter. The “Current Background" level was calculated from the data of the previous section, considering a 7-bin large cluster. 

The most potent improvement is probably the post-process rejection of the background and the treatment of noise. At the moment it is very simplistic and is based on a simple linear background fitting that is not optimal for short acquisition times like in the present test, because the actual number of counts due to cosmic contribution is low and does not average well across the detector strips. One also expects an improvement on the background rejection by optimising the topological selection of tritium-related events. Coincidence muon rejection could also bring the background even closer to zero, strip-wise. An extended area of possible better results exists, on the orange and blue areas of Figure~\ref{Improvement}.

A further step consists in optimizing the detector geometry. At the moment, due to the long unidirectional strips of the reading plane, the background due to noise and cosmic contributions is very high compared to the signal coming from the samples, because the former is integrated over the whole strip length ($\sim$\,12\,cm). As a result, the use of shorter strips or of pads would drastically decrease  the background level. This hypothesis is represented by the "Improved Background" line that was obtained by applying a ratio of 1/12 to the Current Background one. Furthermore, using single pads dedicated to each single cell activity measurement could be beneficial, in the sense that all the events clusters would face a metallic detection area to collect them on the readout plane.

Finally, to improve further the detection sensitivity of the Micromegas-based detector, background could be rejected by classical anti-coincidence techniques 
(red area in Figure~\ref{Improvement}). 

To summarize, several improvement strategies are foreseen to improve the detector. In the case where the tumoral cells activity would be lower than the lowest sample activity considered in this paper (0.3\,Bq), the Micromegas-based detector still has some practical moves to make to perform the task.

\FloatBarrier
\section{Conclusion}
\label{section:Conclusion}
The performance of a new Micromegas-based detector aiming at fulfilling new needs in oncological research  has been tested. Its ability to assess sub-becquerel \tritium~activities has been investigated and compared with a commercial $\beta$-imager. For this, \tritium-glucose was used as a convenient, scalable dummy sample. Activity assessment, linearity and stability from sub-becquerel (0.1\,Bq) to a few tens of becquerels (36\,Bq) per spot have been estimated in a comparative measurement of samples originating from the same \tritium-glucose solutions. \

Overall and in the presence of a broad dynamic range of activities, the Micromegas-based detector proved to do better than its commercial counterpart. 
The observed detection efficiency is constant at 31\% for the Micromegas detector and 23\% (Full dynamic range) and 50\% (Reduced dynamic range) respectively for the $\beta$-imager. Finally, for the present sample size of several hundreds of microns, background treatment (in conjunction with the unfavourable readout plane geometry) sets the difference between the detectors. 
Next steps will concern the measurement of biological tumor samples with an upgraded detector allowing for single cell array deposition with an optimised readout plane. Those smaller samples might separate more thoroughly both detector performances, since the image treatment by the commercial $\beta$-imager might hinder its acquisition process.

\section*{Acknowledgments}
The authors acknowledge the financial support of the Cross-Disciplinary Program on
Instrumentation and Detection of CEA, the French Alternative Energies and Atomic Energy
Commission and the support of the P2I Department of Paris-Saclay University.

\bibliography{mybibfile}

\begin{thebibliography}{1}
\expandafter\ifx\csname url\endcsname\relax
  \def\url#1{\texttt{#1}}\fi
\expandafter\ifx\csname urlprefix\endcsname\relax\def\urlprefix{URL }\fi
\expandafter\ifx\csname href\endcsname\relax
  \def\href#1#2{#2} \def\path#1{#1}\fi

\bibitem{Penner:2012}
N.~Penner, L.~Xu, C.~Prakash,
  \href{https://doi.org/10.1021/tx300050f}{Radiolabeled absorption,
  distribution, metabolism, and excretion studies in drug development: Why,
  when, and how?}, Chemical Research in Toxicology 25~(3) (2012) 513--531,
  pMID: 22309195.
\newblock \href {http://arxiv.org/abs/https://doi.org/10.1021/tx300050f}
  {\path{arXiv:https://doi.org/10.1021/tx300050f}}, \href
  {http://dx.doi.org/10.1021/tx300050f} {\path{doi:10.1021/tx300050f}}.
\newline\urlprefix\url{https://doi.org/10.1021/tx300050f}

\bibitem{Isin:2012}
E.~M. Isin, C.~S. Elmore, G.~N. Nilsson, R.~A. Thompson, L.~Weidolf,
  \href{https://doi.org/10.1021/tx2005212}{Use of radiolabeled compounds in
  drug metabolism and pharmacokinetic studies}, Chemical Research in Toxicology
  25~(3) (2012) 532--542, pMID: 22372867.
\newblock \href {http://arxiv.org/abs/https://doi.org/10.1021/tx2005212}
  {\path{arXiv:https://doi.org/10.1021/tx2005212}}, \href
  {http://dx.doi.org/10.1021/tx2005212} {\path{doi:10.1021/tx2005212}}.
\newline\urlprefix\url{https://doi.org/10.1021/tx2005212}

\bibitem{Lowenthal:1996}
R.~Lowenthal, K.~Eaton, {Toxicity of chemotherapy}, Hematology/Oncolog Clinics
  of North America 10 (4) (1996) 967.
\newblock \href {http://dx.doi.org/https://doi.org/10.1007/s11307-017-1144-0}
  {\path{doi:https://doi.org/10.1007/s11307-017-1144-0}}.

\bibitem{Marusyk:2012}
A.~Marusyk, V.~Almendro, K.~Polyak, {ntra-tumour heterogeneity: a looking glass
  for cancer?}, Nat. Rev. Cancer 12 (2012) 323--334.
\newblock \href {http://dx.doi.org/https://doi.org/10.1038/nrc3261}
  {\path{doi:https://doi.org/10.1038/nrc3261}}.

\bibitem{Giomataris:1995fq}
Y.~Giomataris, P.~Rebourgeard, J.~P. Robert, G.~Charpak, {MICROMEGAS: A High
  granularity position sensitive gaseous detector for high particle flux
  environments}, Nucl. Instrum. Meth. A376 (1996) 29--35.
\newblock \href {http://dx.doi.org/10.1016/0168-9002(96)00175-1}
  {\path{doi:10.1016/0168-9002(96)00175-1}}.

\bibitem{Jambon:2020}
F.~Jambon, E.~Ferrer-Ribas, F.~J.~I. Gutierrez, F.~Beau, V.~Dive, F.~Malloggi,
  A.~Rousselot, F.~Carrel, M.~Trocm{\'{e}},
  \href{https://doi.org/10.1088%2F1742-6596%2F1498%2F1%2F012046}{Medica-plus: a
  novel micromegas detector for high-resolution $beta$ imaging for improved
  pharmacological applications}, Journal of Physics: Conference Series 1498
  (2020) 012046.
\newblock \href {http://dx.doi.org/10.1088/1742-6596/1498/1/012046}
  {\path{doi:10.1088/1742-6596/1498/1/012046}}.
\newline\urlprefix\url{https://doi.org/10.1088%2F1742-6596%2F1498%2F1%2F012046}

\bibitem{Acker2020}
A.~Acker, D.~Atti{\'e}, S.~Aune, J.~Ball, P.~Baron, Q.~Bertrand, D.~Besin,
  T.~Bey, F.~Boss{\`u}, R.~Boudouin, M.~Boyer, G.~Christiaens, P.~Contrepois,
  M.~Defurne, E.~Delagnes, M.~Gar{\c c}on, F.~Georges, J.~Giraud, R.~Granelli,
  N.~Grouas, C.~Lahonde-Hamdoun, T.~Lerch, I.~Mandjavidze, O.~Meunier,
  Y.~Moudden, S.~Procureur, M.~Riallot, F.~Sabati{\'e}, M.~Vandenbroucke,
  E.~Virique, \href{https://hal.archives-ouvertes.fr/hal-02483901}{{The CLAS12
  Micromegas Vertex Tracker}}, {Nucl.Instrum.Meth.A} 957 (2020) 163423.
\newblock \href {http://dx.doi.org/10.1016/j.nima.2020.163423}
  {\path{doi:10.1016/j.nima.2020.163423}}.
\newline\urlprefix\url{https://hal.archives-ouvertes.fr/hal-02483901}

\end{thebibliography}
\bibliographystyle{elsarticle-num}
\end{document}